\documentclass[aps,prb,twocolumn,superscriptaddress,amsmath,amssymb,longbibliography]{revtex4-2}

\usepackage{bm}
\usepackage{amsmath}
\usepackage{amssymb}
\usepackage{graphicx}
\graphicspath{{img_1/}}
\usepackage{color}
\input{epsf}
\usepackage[english]{babel}
\usepackage{gensymb}
\usepackage{ulem}
\usepackage[colorlinks,citecolor=cyan,linkcolor=cyan,urlcolor=cyan,bookmarks=false,hypertexnames=true]{hyperref}

\begin{document}

\title{Skew scattering and ratchet effect in photonic graphene}

\author{O.M.~Bahrova}
\affiliation{Center for Theoretical Physics of Complex Systems, Institute for Basic Science (IBS), Daejeon 34126, Republic of Korea}
\affiliation{B. Verkin Institute for Low Temperature Physics and
Engineering of the National Academy of Sciences of Ukraine, 47
Nauky Ave., Kharkiv 61103, Ukraine}
\affiliation{Institut Pascal, PHOTON-N2, Universit\'e Clermont Auvergne,
CNRS, Clermont INP, F-63000 Clermont-Ferrand, France}
\author{S.V.~Koniakhin}
\affiliation{Center for Theoretical Physics of Complex Systems, Institute for Basic Science (IBS), Daejeon 34126, Republic of Korea}
\affiliation{Basic Science Program, Korea University of Science and Technology (UST), Daejeon 34113, Republic of Korea}

\begin{abstract}



The present paper is devoted to a comprehensive theoretical study of asymmetric (skew) scattering in photonic graphene, with the main focus on its realization with semiconductor microcavity exciton-polaritons. As an important consequence of the skew scattering, we prove the appearance of the ratchet effect in this system. Triangular defects in the form of missing micropillars in a regular honeycomb lattice are considered as ones that break the spatial inversion symmetry, thus providing the possibility of the ratchet effect. By means of the numerical solution of the effective Schr\"odinger equation, we provide microscopical insight into the process of skew scattering and determine indicatrices, cross-sections, and asymmetry parameters. In a system with multiple coherently oriented triangular defects, a macroscopic ratchet effect occurs as a unidirectional flux upon noise-like initial conditions. Our study broadens the concept of ratchet phenomena in the field of photonics and optics of exciton-polaritons.

\end{abstract}
\maketitle

\section{Introduction}\label{sec:intro}



In the domain of condensed matter physics, the study of exciton-polaritons has emerged as a captivating direction, offering a unique blend of quantum optics and solid-state physics. Exciton-polaritons, including them existing in semiconductor microcavities, are hybrid quasiparticles arising from the strong coupling~\cite{Weisbuch1992} of 2D confined excitons and long-living photons in specially engineered optical cavities~\cite{Malpuech2003}. Due to composite nature sharing properties of light and matter~\cite{Solnyshkov_2021a}, multiple non-trivial phenomena were observed in systems of exciton-polaritons~\cite{Boulier2020,Bloch2022}. Often these phenomena have analogues in other condensed matter systems. For example, polaritons manifest possibility to condensate at high temperatures~\cite{Kasprzak2006}. The effects of superfluidity~\cite{Amo2009} and soliton emergence allow attributing polariton systems as ``quantum fluids of light''~\cite{Carusotto2013, Byrnes2014,Solnyshkov2014}. Abrikosov vortices have their counterparts in polariton systems~\cite{Lagoudakis2008,boulier2015vortex}, and the complex phenomenon of quantum turbulence~\cite{Berloff2010,Reeves2012,Koniakhin2020} is also of high interest in that context. Exciton-polaritons offer flexible all-optical control of excitation~\cite{Wertz2010} and various detection techniques for cw and time-resolved measurements of photon emission spatial profile, momentum-domain emission and energy dispersion. As a result, all mentioned above phenomena can be experimentally investigated with an unprecedented level of details and entirety. The artificial optical lattices obtained by advanced microcavity etching techniques are of particular interest. 


Photonic lattices have gained significant attention due to the ability to both mimic certain phenomena observed in real 2D crystals and practically implement theoretical concepts lacking analogues in nature. A site in such structures is a single micropillar~\cite{Ferrier_2011} and further these micropillars can be organized into complicated arrays~\cite{galbiati2012polariton,Solnyshkov_2016,harder2021coherent}. Similarly to electronic band structures in graphene~\cite{wallace1947band}, photonic honeycomb lattices exhibit the formation of Dirac cones in the band structure and the associated linear dispersion relation~\cite{Nalitov_2015, Oudich2021}. Thus there are extensive studies of the topological and transport properties of artificially created graphene-type lattices~\cite{Zhang_2022,Solnyshkov_2018,Solnyshkov_2016,Bahrova_2024}.
In general, topological photonics is among hot topics of modern physics nowadays. It opens a path towards the investigation of fundamentally new states of light and potential applications of topological phenomena in the development of robust and scalable quantum devices~\cite{Gregersen2017,Kavokin2022,Ricco2024}.
Manifestation of topologically protected edge states and the associated quantum Hall-like effects~\cite{Nalitov_2015,Solnyshkov_2016,Solnyshkov_2016a} in microcavity polariton lattices allows them to gain significant attention. The closely related phenomenon is the Hall effect (both in optical and electronic systems including the variations like spin Hall effect, valley Hall effect etc.). Its origin can lie in the manifestation of geometric phase or in the effect of asymmetric scattering. 

Asymmetric (i.e., skew) scattering in physics refers to a type of scattering process where the probabilities of scattering into different directions are not symmetrical with respect to the incident direction~\cite{Belinicher_1980}. Thus it introduces directionality which can arise from structural asymmetries, non-uniform potentials, or the presence of external fields, etc., into scattering processes, where certain directions become favored or disfavored based on the system's characteristics. This phenomenon can occur in various physical systems~\cite{Drexler_2013,Gregersen2017,Gregersen2018} and understanding asymmetric scattering is crucial for interpreting experimental results and designing devices with specific functionalities.

Furthermore,  inversion symmetry breaking, including one originated from defects of particular symmetry, in electronic or mechanical systems can lead to the so-called ratchet effect. The ratchet effect itself can be referred as appearance of the steady unidirectional response to an oscillating or stochastic driving excitation~\cite{Belinicher_1980}. 
The geometry of ratchet effect is defined by global symmetry of the system and for the relevant case of C$_{3v}$ symmetry, one can expect appearance of vertical response component for driving force applied horizontally~\cite{Koniakhin_2014}.

In graphene under asymmetric periodic strain the classical ratchet effect was considered~\cite{Nalitov_2012}, see also Refs.~\cite{Monch_2022,Olbrich2016}. Moreover, the directed electron motion in a two-dimensional electron gas~\cite{Budkin2016,sassine2008experimental} and single graphene layer~\cite{Drexler_2013,Bellucci2016} which is subject to a static magnetic and alternating electric fields, was observed. Asymmetry in the geometry of a microstructure can be a basis to implement the diode device~\cite{mencarelli2022current} which is closely related to the field of ratchet studies.

As mentioned above, asymmetric (skew) scattering is one of the possible underlying mechanisms for the realization of ratchet effect. Thus the skew scattering on semidisk Galton board~\cite{chepelianskii2007photogalvanic,ermann2011relativistic,sassine2008experimental} and triangular defects~\cite{Koniakhin_2014} was proven to initiate ratchet effect in low-dimensional structures and graphene, see also Ref.~\cite{Hild2023}.

However, within the framework of ratchet effect, photonic systems differ significantly from electronic ones. In numerous electronic systems~\cite{Koniakhin_2014,Nalitov_2012}, one deals with fixed Fermi level defining the carrier density and irradiation playing a role of external periodic driving force. On the contrary, in photonics, laser pumping serves as a source of particles and its effect on the potential experienced by the particles and thus on their spatiotemporal dynamics is less pronounced. Therefore, to study ratchet effect in exciton-polariton systems, one should engineer alternative excitation schemes \textit{beyond} the excitation by unstructured laser light (either in quasi-resonant or non-resonant regime for polaritons) which would still be effective for electronic systems. At the same time, due to the mentioned above flexibility in all-optical excitation and wave function detection techniques, photonic systems are essential to simulate and visualize the ratchet effect microscopic mechanisms, whose direct observation is inaccessible in electronic systems.


In the present paper we demonstrate a possibility to observe ratchet effect in photonic graphene as the consequence of skew scattering on triangular C$_{3v}$ symmetry defects embedded into a regular honeycomb lattice.
The remainder of this paper is organized as follows. In the beginning of Sec.~\ref{sec:1tr} the model system based on a polariton analog of graphene with triangle-shaped defects is introduced. 
Then by means of the effective Schr\"odinger equation (SE) for exciton-polaritons, we study in details properties of the asymmetry in scattering. We derive scattering cross-sections and indicatrices for the wave packet (with a specific wave vector) that has been scattered on a single triangular defect in classical and wave regimes. Further, in Sec.~\ref{sec:cluster} we explore setup of the honeycomb lattice with multiple randomly distributed defects and provide results of the corresponding numerical simulations. We propose a simple scheme to observe the emergence of ratchet effect based on the platform of exciton-polaritons with real-space time-integrated intensity measurements. In Sec.~\ref{sec:kinetics} we establish connection between the results from previous sections with that obtained employing Monte-Carlo simulations being a simple realization of the kinetic approach. Finally, in Sec.~\ref{sec:conclusions} we conclude the paper with a summary and an outlook.

\section{Skew scattering by single triangle}\label{sec:1tr}

\begin{figure}
    \begin{center}
        \includegraphics[width=\linewidth]{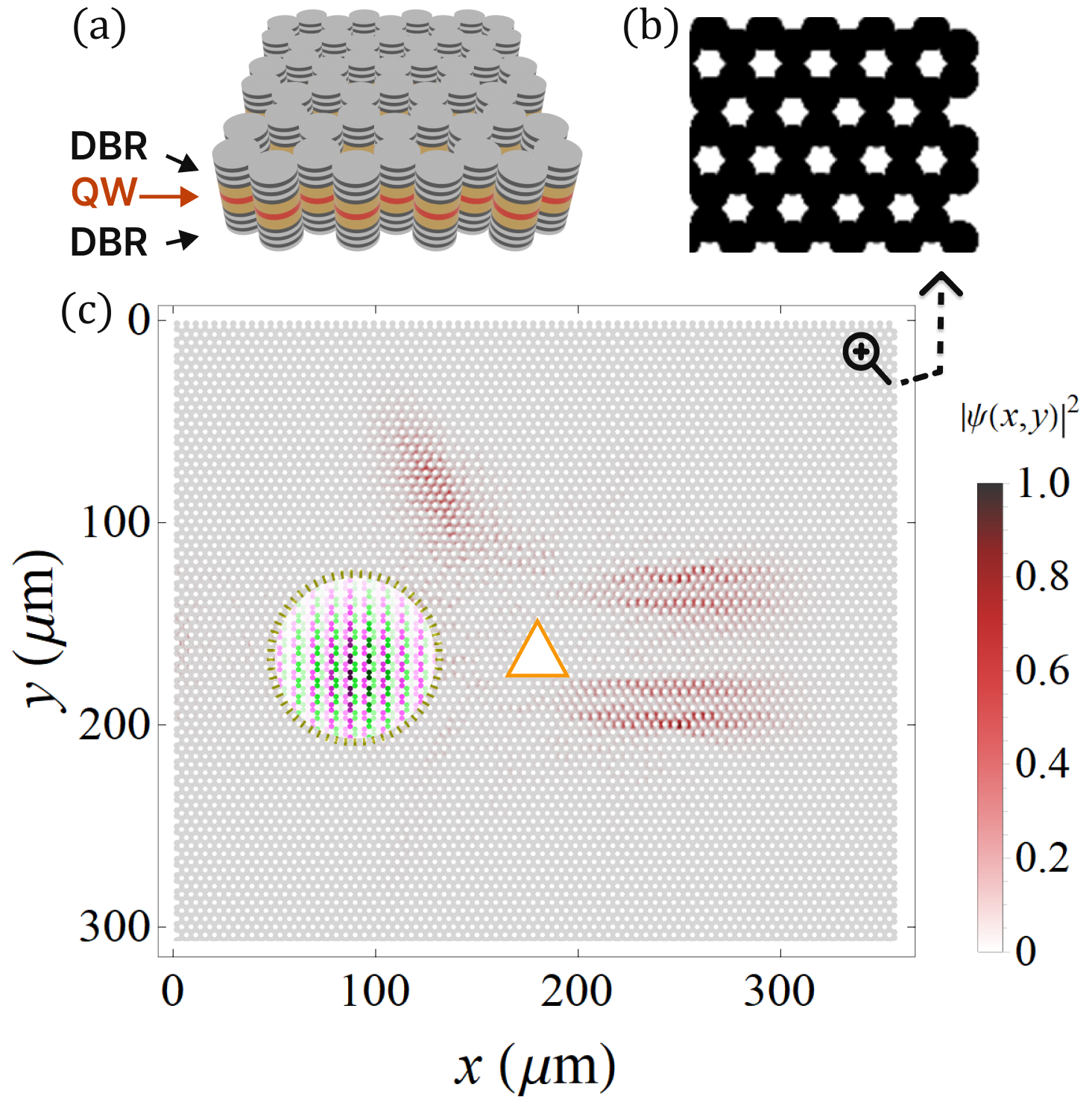}
    \end{center}
    
    \caption{Demonstration of asymmetric Gaussian wave packet scattering on triangular defect (orange triangle) embedded into polariton graphene lattice (shown in gray in the main figure). The yellow circle indicates the half-width of the initial Gaussian-shape wave function. Green-purple colors with intensity modulation show phase profile of the wave packet for the wave vector $q=0.2$ (see text), the color intensity is proportional to the wave function amplitude squared.
    }
    \label{fig:fig1}
\end{figure}

In this work we consider an optical exciton-polariton analog of a single-layer graphene. That system has its experimental realization via the microcavity etching technique resulting in the overlapped micropillars arranged into honeycomb lattice~\cite{Real_2020,Jamadi_2020}. We focus on the ratchet effect originated from skew scattering by the defects of particular shape. Specifically, equilateral triangle-shaped defects (C$_{3v}$ symmetry) are embedded into graphene lattice as missing micropillars (sites), see Fig.~\ref{fig:fig1}. Formally, presence of such defects breaks the spatial inversion symmetry of the system and provides the global C$_{3v}$ symmetry.

We investigate time evolution of a wave packet (or initial wave function engineered in differing way) in linear regime and perform numerical simulations of the effective Schr\"odinger equation widely used to trace dynamics of exciton-polaritons at low densities:
\begin{equation}\label{SE}
i\hbar \frac{\partial\psi }{\partial t} = \left[ -\frac{\hbar^2}{2m}\left(\partial_x^2+\partial_y^2\right) + V  - 
i\hbar\frac{\gamma}{2}\right] \psi  . 
\end{equation}
Here $m=5\cdot 10^{-5}~m_e$ is the effective polariton mass, $\gamma\equiv\gamma_0 f(x,y)$ stands for the polariton decay rate, $V(x,y)$ describes the confinement potential (3~meV for the barrier height) defined by the shape of honeycomb lattice of micropillars with embedded triangular defects as it is shown in Fig.~\ref{fig:fig1}. We perform integration of Eq.~(\ref{SE}) on a $2048 \times 2048$ square mesh grid with $0.2~\mu$m spatial step using the third-order Adams-Bashforth method with time step $\delta t =2\times 10^{-5}$~ps and putting into service Graphics Processing Unit (GPU) for acceleration. The pillar size, $d_0=3.2~\mu$m, corresponds to the one in experimental samples from Refs.~\cite{Jamadi_2020}. At the same time, the obtained results are valid for the optical realization of graphene by means of electromagnetic induced transparency (EIT) effect in atomic vapor cells~\cite{zhang2019particlelike,Zhang_2022,li2023simultaneous}.

As a first step, we investigate properties of scattering on a single triangular defect. Thus, initial conditions are a plane wave modulated by a Gaussian function as follows:
\begin{equation}\label{psi_in}
    \psi_{in} = \frac{1}{N} \exp\left[ -\frac{(x-x_0)^2}{2R_0^2}-\frac{(y-y_0)^2}{2R_0^2} + \mathbf{kr}\right],
\end{equation}
with the wave vector values $\mathbf{k}=(k_x,k_y)$ taken in the proximity of Dirac points within the first Brillouin zone: 
$\mathbf{k}=\mathbf{k_D}-|\mathbf{k_D}|\mathbf{q}$, where $\mathbf{k_D}$ stands for the wave vector of a chosen Dirac point and $\mathbf{q}$ defines the wave vector components referenced from that Dirac point. It is useful to keep the latter dimensionless and normalize to  $\vert\mathbf{k_D}\vert$.

\begin{figure}
    \centering
    \includegraphics[width=0.99\linewidth]{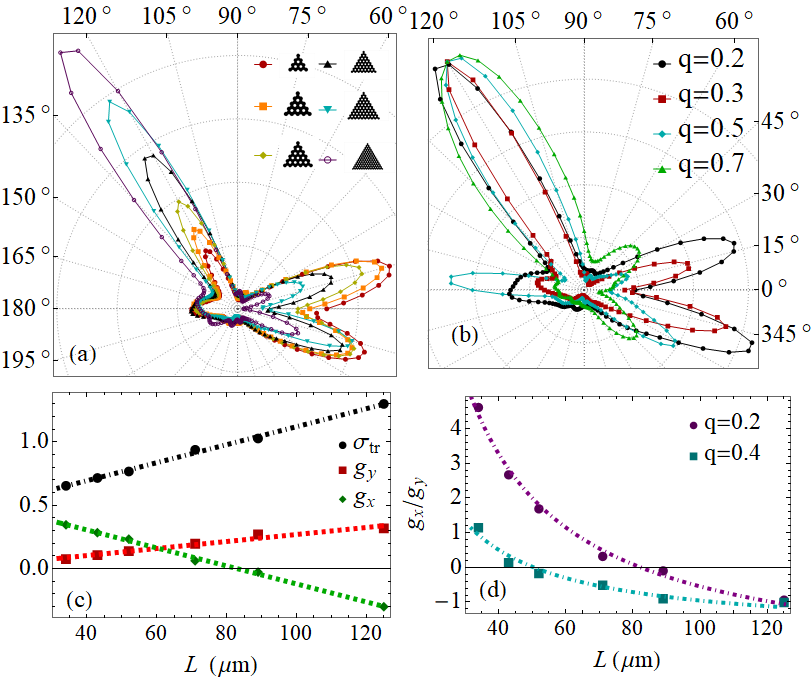}
    \caption{Scattering indicatrices for triangular defect of various sizes (a) and values of the initial wave vector (b). The black curve is common in panels (a) and (b). Panel (c) shows the coefficients of scattering asymmetry $g_{x,y}$ along horizontal $x$ and vertical $y$ axes (red and green markers, respectively), given by Eq.~(\ref{coeffg}), for $q=0.2$ and compared with the transport cross-section $\sigma_{tr}$ (black dots). 
    The dashed lines mark the linear fit. Panel (d) represents the ratio of $g_x/g_y$ versus size of the triangle defect $L$. 
    }
    \label{fig:fig2}
\end{figure}
One can characterize a scattering process via determination of the differential cross-section specifying angular distribution of the scattered wave function. In Figure~\ref{fig:fig2} one can see calculated scattering indicatrices $W_{0\degree ,\varphi'}$ that yield the fraction of optical signal ($|\psi|^2$) scattered in a given direction $\varphi'$ for the $0^\circ$ incident direction as it is presented in Fig.~\ref{fig:fig1}. To derive the indicatrix magnitude at a given angle, we integrate in real space signal intensity in the corresponding plane angle with the defect center as an origin point. In experiment this procedure can be reproduced by the near-field photoluminescence measurement. We take wave packet size $R_0 = 90$~$\mu$m to be greater than the wavelength $\lambda$. We take into account defects of different sizes covering various regimes of wave vector scattering: from wave to classical. 
The initial wave vector has its component only in the $x$ direction with the value $q\equiv \vert \mathbf{q}\vert=0.2$ measured from the Dirac point $\mathbf{k_D}=(2\pi/(3\sqrt{3}a),2\pi/(3a))$ with $a$ being the lattice constant.

The following hierarchy of the size scale takes place in the setup under consideration. In \textit{wave} regime $R_0 \gg L \gtrsim \lambda$, the wave packet size is greater than defect size $L$ and the latter is in its turn greater than wave length, $\lambda = \frac{2\pi}{k}$. This regime can also be  referred as quantum mechanical because the spatial structure of wave function and scatterer potential, as well as their interplay, are of importance. In optics it corresponds to Mie scattering. By increasing size of the defect, one can approach the regime $L \gtrsim R_0 \gg \lambda$, where the wave packet behaves like a \textit{classical} point-like particle undergoing reflection by a wall. In both cases, wave length is the lowest size scale. In the case of small defects $L\ll \lambda\ll R_0$ (Rayeigh limit in optics), scattering efficiency, including the asymmetric part, becomes too low to have manifestation in detectable ratchet flux. Figure~\ref{fig:fig2} shows that side lobes in the scattering indicatrices responsible for asymmetric scattering and ratchet effect grow monotonically as a function of $qL$, which qualitatively agrees with findings of Ref.~\cite{Koniakhin_2014}, however direct comparison of these results is ambiguous due to differing regimes (in particular, infinite potential of defect here) and small defect size range covered in present simulations due to physical system size limitations. The most robust correspondence takes place deeply in the classical limit with linear dependence in defect size, $L$, and irrelevance of the wave vector. The transmitted signal vanishes and the scattering indicatrix becomes $\delta$-function shaped with intuitive wall reflection rules, see Fig.~S2(b) in Supplemental Material~\cite{supp}.



\begin{figure}
    \centering
    \includegraphics[width=0.99\linewidth]{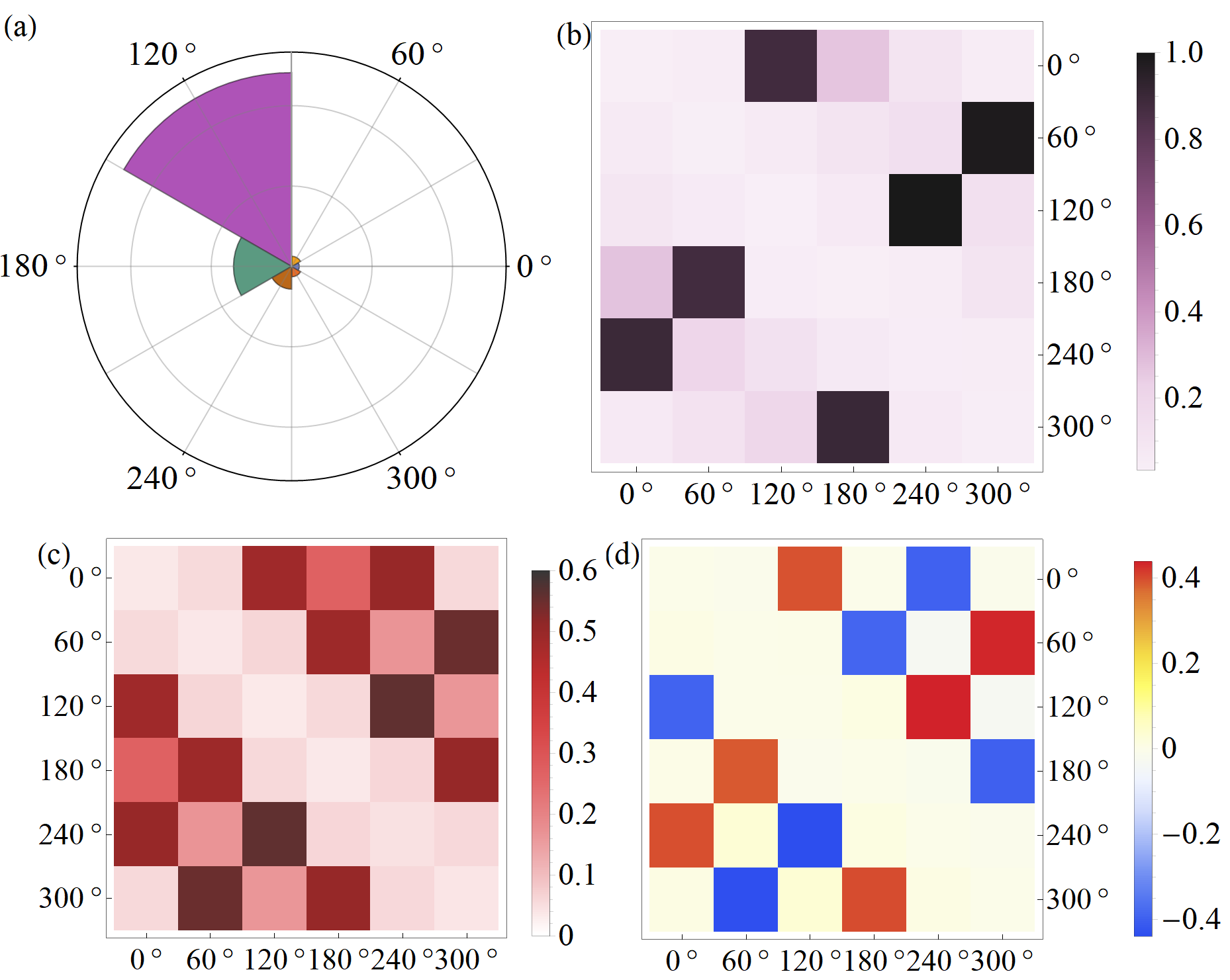}
    \caption{Panel (a). Discretized scattering indicatrix for scattering process with geometry as in Fig.~\ref{fig:fig2} in classical regime $L>R_0$ (defect is larger than the wave packet size). Panel (b). Scattering rate $W_{\mathbf{k},\mathbf{k}'}$ matrix for six discrete directions of wave vector. Vertical axis is for incident direction $\mathbf{k}$ and horizontal axis is for scattered signal direction $\mathbf{k}'$. Panels (c) and (d) show decomposition of scattering matrix into symmetric ($W_{\mathbf{k},\mathbf{k}'}^{(s)}$) and anti-symmetric ($W_{\mathbf{k},\mathbf{k}'}^{(as)}$ ) parts, respectively. Noteworthy, panel (a) in fact represents the values $W_{0,\mathbf{k}'}$.}
    \label{fig:fig3}
\end{figure}

To focus on the scattering process itself, we take $\gamma_0=0$ thus assuming the polariton lifetime being much longer than the time required for the scattering process to occur. This assumption should also be valid for EIT-based realizations of photonic graphene in atomic vapor cells~\cite{Zhang_2022}.
Then, in order to characterize the asymmetry of scattering, we introduce the asymmetry parameter~\cite{Bohren1983}  $\mathbf{g}\equiv (g_x,g_y)$ defined as average cosine (sine) of scattering angle: $(g_x,g_y)= \left(\langle \cos{\varphi}\rangle ,\langle \sin{\varphi}\rangle \right)$, e.g.:
\begin{equation}\label{coeffg}
    g_{y}=\frac{1}{w}\int_0^{2\pi}  \sin{\varphi} W_{0\degree,\varphi}d\varphi ,
\end{equation}
which quantify asymmetry along horizontal ($g_x$) and vertical ($g_y$) axes; 
$w=\int_0^{2\pi}  W_{0\degree,\varphi}d\varphi$.
Thus, using this approach, we become capable to classify polar scattering diagrams as shown in Fig.~\ref{fig:fig2}(c), where the component dependencies of the asymmetry vector on the defect size are plotted (red and green markers, respectively). For comparison, we also present by the black dots data for transport (momentum transfer) cross-section $\sigma_{\rm tr}=\langle 1-\cos{\varphi}\rangle$ defined in accordance with Eq.~\ref{coeffg}. The dashed lines indicate linear fitting drawn as the guide for an eye. The linear dependence of asymmetry in scattering versus triangle size has been predicted for regular graphene~\cite{Koniakhin_2014}. Note that $g_{x(y)}$ takes values in the range $[-1,1]$ by definition, 
$\mathbf{g}=(1,0)$ corresponds to the limit of absence of a scatterer. 
Apparently, Fig.~\ref{fig:fig2}(c) shows that $g_y$ is always positive which reflects the fact of prevailing of scattering in the upper semiplane, see Fig.~\ref{fig:fig2}(a). Moreover, $g_x>0$ is related to the feature when a defect scatters more wave function toward the forward direction ($\varphi'=0$), and vice versa, $g_x<0$ when the back-scattering ($\varphi'=\pi$) prevails. According to $g_x$ and $g_y$ behaviors, the transition from the wave to classical mirror reflection regime takes place at approximately $90$~nm  which matches with high precision with the wave packet size. Finally, figure~\ref{fig:fig2}(d) shows the ratio of $g_x/g_y$ versus size of triangle $L$ for two values of initial wave vector $q=0.2$ (purple dots, associated with (a) and (c) subfigures) and $q=0.4$ (rectangles), see also Supplemental Material~\cite{supp}.

From the physical meaning of the asymmetry parameter, in the vanishing defect limit $L\rightarrow0$, one obtains $g_x/g_y\rightarrow 1/0 = \infty$ since we have only horizontally transmitted signal and no deviation along y-axis. In the classical limit, the trajectory of wave packet after scattering is defined with $\delta$-function scattering rate $W$. As a result, $g_x/g_y \rightarrow\mathrm{const}$. For sufficiently large defect size $L \gg R_0$, the limit value is $-1/\sqrt{3}$, and in the plots we see the transitional regime.

\begin{figure}
    \centering
    \includegraphics[width=0.99\linewidth]{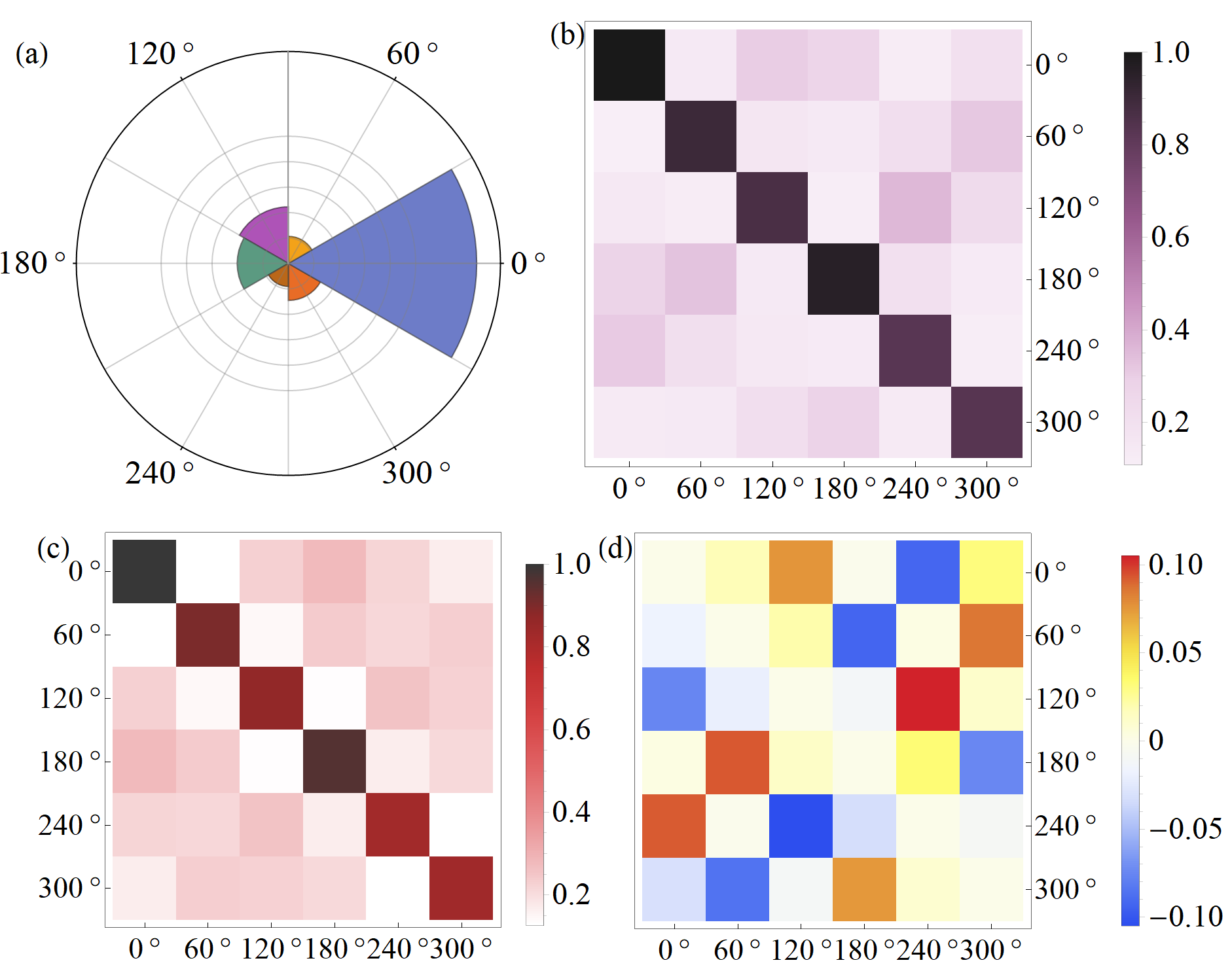}
    \caption{Scattering diagram and scattering rate matrices for the wave regime~$R_0 \gg L$.}
    \label{fig:fig4}
\end{figure}

As a next step, it becomes essential to construct full scattering rate matrices for different defect size regimes and various incident and outgoing directions. We discretize wave vector directions with 60$\degree$ step. In the classical limit only these 6 directions are allowed if wave packet moves initially parallel to triangles' sides. In the wave limit with sufficiently continuous scattering diagrams, see Fig.~\ref{fig:fig2} (a),(b), we integrate scattered signal in the given plane angle around central values indicated in Figs.~\ref{fig:fig3} and~\ref{fig:fig4} (b)-(d). The chosen discretization is in agreement  with defect symmetry $C_{3v}$ (and also with graphene crystal symmetry $C_{6v}$) thus allowing constructing hereinafter reasonable kinetic-like Monte-Carlo simulations.
%
The scattering rate $W_{\mathbf{k},\mathbf{k}'}$ defines probability distribution for an incident direction $\mathbf{k}$ to one of scattering $\mathbf{k}'$ \ and is represented as $6 \times 6$ matrix. Furthermore, following the definition given by Eq.~(4.2) in Ref.~\cite{Belinicher_1980} and taking into account discreteness of wave vector directions, we decompose scattering rate matrices into symmetric $W^{(s)}$ and anti-symmetric $W^{(as)}_{\mathbf{k},\mathbf{k}'}$ parts as $W^{s(as)}_{\mathbf{k},\mathbf{k}'}=\left(W_{\mathbf{k},\mathbf{k}'}\pm W_{\mathbf{k}',\mathbf{k}}\right)/2$. 
 This allows one to map the hallmarks of asymmetry in wave function scattering on the characterization of a square matrix, see Sec.~\ref{sec:kinetics} for more detailed consideration.
Figures~\ref{fig:fig3}-\ref{fig:fig4} show scattering rate matrices with its symmetric and anti-symmetric parts in two opposite regimes of scattering: classical and wave, respectively, see also Supplemental Material~\cite{supp}. It is worth to mention that anti-symmetric part of the scattering matrix allows us to note differences in magnitudes and directions (represented by the sign) for skew scattering in respect to a symmetric case. One could make a correspondence between Fig.~\ref{fig:fig3}(a) and the purple curve as well as Fig.~\ref{fig:fig4}(a) with the red curve in Fig.~\ref{fig:fig2}(a). 

\section{Ratchet effect}\label{sec:cluster}

\begin{figure}
    \centering
    \includegraphics[width=0.99\linewidth]{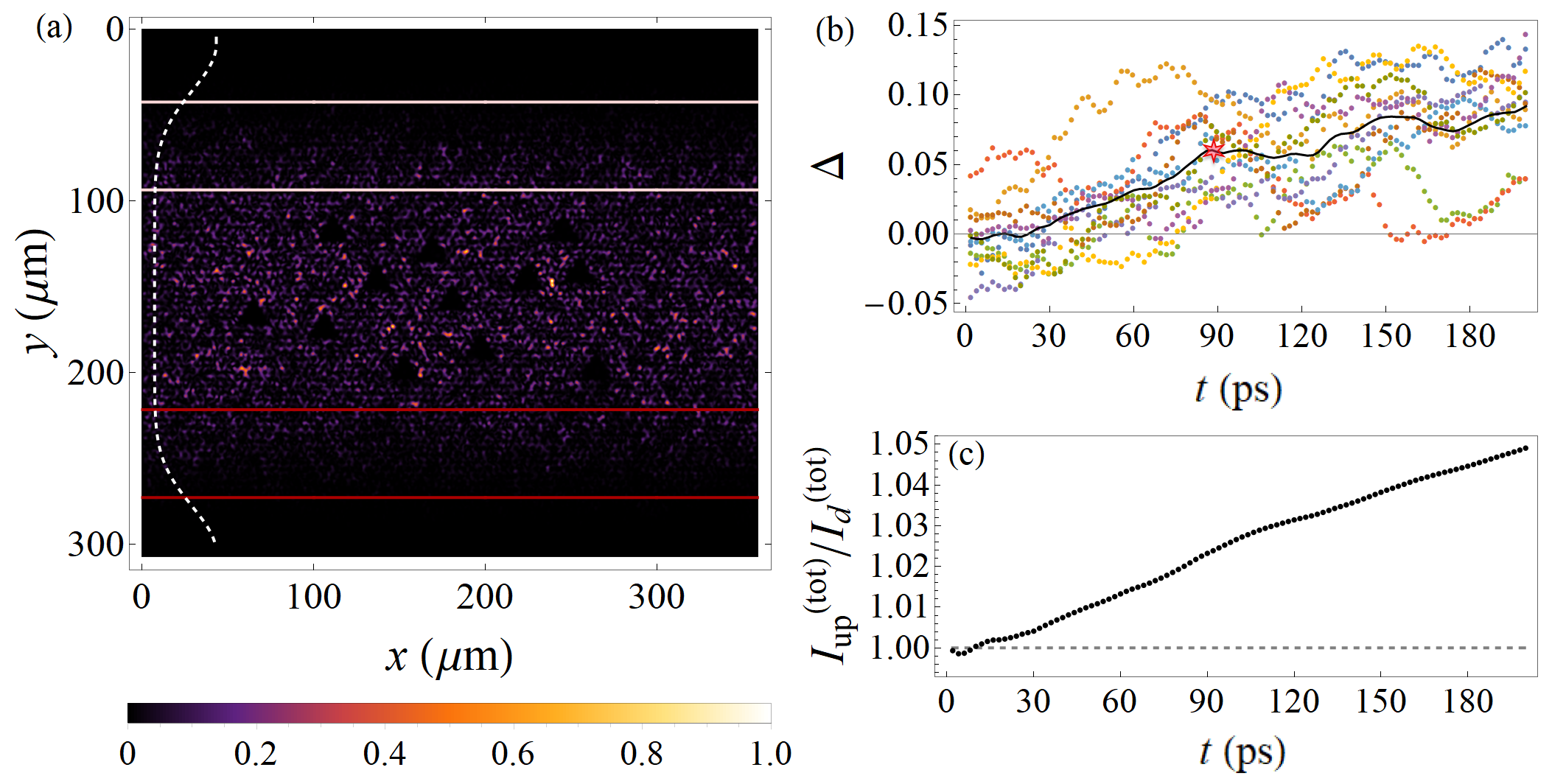}
    \caption{Illustration of macroscopic ratchet effect in the system. (a) The wave function $|\psi (x,y)|^2$ for $t=90ps$ (marked by the red star in (b)). The light-red and dark-red lines show the wave function counting regions for comparison of upwards and downwards fluxes, respectively. The white dashed curve indicates the spatial decay rate profile $\gamma (y)$. (b) The difference $\Delta$ of total emission intensity (proportional to polariton density $|\psi|^2$) from upper (up) and lower (d) counting regions over time for various initial wave functions (in colors) and the average value (black curve). (c) The ratio of time-averaged wave function in upper region $I_{\mathrm{up}}^{\mathrm{(tot)}}$ to the one in lower counting region $I_{\mathrm{d}}^{\mathrm{(tot)}}$ versus time (averaged over 10 realizations as showed in (b)).}
    \label{fig:fig5}
\end{figure}

In the present section we demonstrate how skew scattering on triangular defects leads to ratchet effect in the whole system and quantitatively characterize it. We consider an array of randomly distributed defects between upped and lower absorbers characterized by finite polariton losses $\gamma (x,y)$. Importantly, despite random positions, the defects are coherently oriented, which provides the global C$_{3v}$ symmetry for the system required to observe the ratchet effect. The absorbers exclude effects of scattering from the boundaries. The employed decay rate spatial profile is homogeneous in horizontal direction and its vertical profile $\gamma (x,y)\equiv\gamma (y)$ is shown in Fig.~\ref{fig:fig5}(a) with the white dashed curve. Maximal magnitude of the decay rate equals $\gamma_0=10$~ps$^{-1}$. We start from a white noise-like wave function, i.e., the whole available lattice between the detection regions is populated with random phase and amplitude at each site  (see Supplemental Material~\cite{supp}) to demonstrate resulting unidirectional response. As it was underscored above, optical systems sufficiently differ from electronic ones with respect to excitation techniques and it is challenging to experimentally apply a potential equivalent to oscillating or stochastic electric field accelerating electrons. Present mode of excitation (realized via e.g. quasi-resonant pumping for polaritons) can be treated as excitation of multiple harmonics with random amplitudes and phases.

In order to quantitatively characterize manifestation of the ratchet effect, we integrate over time the wave function intensity ($\vert \psi\vert ^2$) in the upper and lower counting regions denoted by pink and red lines in Fig.~\ref{fig:fig5}(a) and placed slightly prior to the absorbers. The counting regions are symmetric with respect to a perfect structure horizontal symmetry axis and to the absorbers. To exclude the effect of statistical asymmetry, the triangular defects are placed in equivalent fractions above and below the symmetry axis. Figure~\ref{fig:fig5}(a) shows the intensity profile of wave function after 90~ps of time evolution.

In Figure~\ref{fig:fig5}(b) we plot the instantaneous characteristic of asymmetry in resulting ratchet flux, i.e. difference in the intensities for the upper and the lower counting regions at a given time moment $t$:
\begin{equation}
    \Delta (t) = \frac{I_{\mathrm{up}}(t)-I_\mathrm{d}(t)}{0.5\cdot(I_{\mathrm{up}}(t)+I_\mathrm{d}(t))}, 
    \end{equation}
where $I_{\mathrm{up(d)}}(t)=\int_{\mathrm{up(d)}}\vert \psi (x,y,t)\vert^2 dxdy$, for different realizations (colors) of initial conditions for the wave function. The black dots correspond to the averaged values.
Further, Fig~\ref{fig:fig5}(c) shows the integral characteristic being the ratio of time-averaged intensities between the upper (up) and the lower (d) counting regions $$I_{\mathrm{up(d)}}^{\mathrm{(tot)}}(t)=\frac{1}{t}\int_0^{t} I_{\mathrm{up(d)}}(t') dt'.$$

Experimentally, observation of ratchet effect will require growing and etching the structures several times larger than modern ones~\cite{Jamadi_2020}. The increase of lifetime will be also sufficiently beneficial for manifestation of the ratchet effect within the proposed setup. The creation of the ``noise''-like wave function can be possible by using spatial light modulators. However, experimental investigation of skew scattering by a triangular defect is accessible by etching appropriate lattice in contemporary samples. The pulsed quasi-resonant excitation of wave packet combined with time-delayed near-field photoluminescence will be sufficient for restoration of scattering indicatrices and diagrams.


\section{Kinetic approach to ratchet effect}\label{sec:kinetics}

In present section, we aim to establish connection between the derived scattering cross sections (Sec.~\ref{sec:1tr}) for single triangle and observed macroscopic ratchet effect in the system (Sec.~\ref{sec:cluster}) by means of a statistical approach and Monte-Carlo simulations.


To do so, we implement the simulation of carrier random walks with six discrete directions of motion corresponding to the angles in panels (a) of Figs.~\ref{fig:fig3}-\ref{fig:fig4}. At each time step the particle might move at distance $dl$ or undergo scattering process with a small probability $p_{dl} = n\sigma dl$. The integral cross-section can be estimated as triangle size: $\sigma \approx L$. Scattering probabilities for various incident and resulting angles are encoded in (b) panels of Figs.~\ref{fig:fig3}-\ref{fig:fig4}. Finally, the scattering rate is proportional to the concentration of defects (triangles) $n$ taken as in the setup used for the Schr\"odinger equation simulations discussed above. The initial position of the particle $(x_0,y_0)$ is chosen to be arbitrary within the sample size $L_0$ and motion direction is taken randomly from 6 allowed values, which mimics the noise-like wave function used in Schr\"odinger equation simulations.



\begin{figure}
    \centering
    \includegraphics[width=0.99\linewidth]{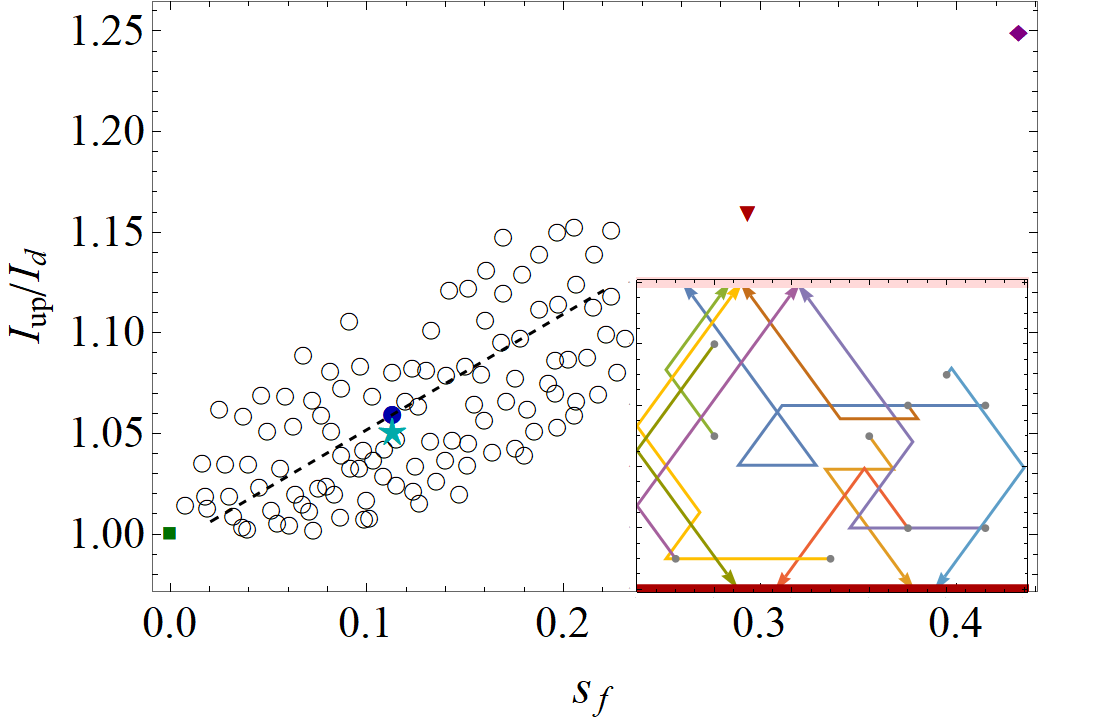}
    \caption{Illustration of the emergence of ratchet effect due to asymmetry in scattering via stochastic random-walk approach. The vertical axis shows ratio of upwards to downwards carrier fluxes. The horizontal axis is for a measure of scattering asymmetry. Blue star corresponds to the value obtained within the SE simulations. Other colored markers represent results of Monte-Carlo simulations with scattering rates defined by the ones from full Schr\"odinger equation calculations. The empty circles are related to values of ratchet effect obtained with synthetic matrices (see the main text). The inset shows examples of the conspicuous particle's trajectories.  }
    \label{fig:fig6}
\end{figure}

We have run sets of multitudinous ($\propto 10^6$) independent Monte-Carlo simulations of random walks and detected whether trajectory of the particle ended up at the lower or at the upper boundary. The ratio between the number of cases when the particle stops on the upper boundary ($y_f=L_0$) and vice versa ($y_f=0$), $I_{\mathrm{up}}/I_{\mathrm{d}}$, is calculated and plotted in Fig.~\ref{fig:fig6} (examples of particle's trajectories are shown in the inset). In Figure~\ref{fig:fig6} results for three previously considered types of scatterers
are indicated by the solid markers: blue disk (wave regime), red triangle (intermediate), and purple rhombus (classical). In addition, the cyan star gives the value of ratchet effect that was obtained within full Schr\"odinger equation simulations, see Fig.~\ref{fig:fig5}, in order to emphasize self-consistency of the approaches.
In Figure~\ref{fig:fig6}, we also indicate by the green square marker the case of an uniform probability distribution which describes the scattering on a $C_{6v}$-symmetric defect and therefore absence of asymmetry in scattering.


One can note that the general structure and dominating elements of the asymmetric parts for all considered scattering matrices are similar, see Sec.~\ref{sec:1tr} and Supplemental Material~\cite{supp}. As a result, it is possible to characterize  scattering matrix asymmetry by a single value as
\begin{equation}\label{sf}
    s_f=\frac{\| A^{as} \|_F}{\| A^{s}\|_F +\| A^{as} \|_F},
\end{equation}
Here $A^{as(s)}=\left(A\mp A^T\right)/2$ is the anti-symmetric (symmetric) part of scattering matrix $A$ and $\|\cdot \|_F$ denotes the Frobenius norm of a matrix (in principle, any matrix norm works). The $s_f$ values calculated for the scattering matrices of the treated defects, give the horizontal axis values in Fig.~\ref{fig:fig6}. 

Further, one can construct a synthetic scattering matrix of arbitrary degree of asymmetry based on the one corresponding to, e.g., to the wave regime, see Fig.~\ref{fig:fig4}(b). Using the decomposition of a matrix into symmetric and anti-symmetric parts, one can utilize the following expression: 
\begin{equation}\label{tildaW}
    \tilde{W}_{\mathbf{k},\mathbf{k}'}=a_1 W_{\mathbf{k},\mathbf{k}'}^{(as)}+a_2 \left(W_{\mathbf{k},\mathbf{k}'}^{(s)}-W_{\mathbf{k},\mathbf{k}}^{(s)}\right)+a_0 W_{\mathbf{k},\mathbf{k}}^{(s)},
\end{equation}
with a definite value of asymmetry $s_f$ adjusted by the coefficients $a_i, i=1,2,0$. Here, $a_{1}$ controls the degree of asymmetry, $a_2$ gives the contribution from the off-diagonal symmetric part elements and $a_0$ is for the diagonal part. The normalization condition $\sum_{\mathbf{k}'}\tilde{W}_{\mathbf{k},\mathbf{k}'}=1$ is also satisfied by tuning the amplitude of the coefficients. Noteworthy, various ratios between the coefficients can give the same measure of asymmetry $s_f$. Results of the Monte-Carlo simulations for multiple synthetic scattering matrices are presented by open circles in Fig.~\ref{fig:fig6}. One can note from Fig.~\ref{fig:fig6} that ratchet current quantified by $I_{\mathrm{up}}/I_{\mathrm{d}}$ and emerged due to asymmetry in scattering processes, is linearly proportional to the asymmetry coefficient, $s_f$. This fact is in agreement with the predictions of Ref.~\cite{Belinicher_1980} for the flux of noninteracting particles experienced elastic scattering in the medium which does not posses a center of symmetry. Additionally, one could note that the coefficient of asymmetry is, in turn, proportional to the characteristic size of the triangle: $s_f\propto L$. In agreement with Ref.~\cite{Koniakhin_2014}, in the classical regime (i.e. at higher $s_f$), asymmetric part of scattering rate is dominating and thus the ratchet flux is expected to be much stronger.

%

\section{conclusions}\label{sec:conclusions}


We consider a way to reveal the presence of ratchet effect in the exciton-polariton analogue of graphene employing the phenomenon of skew (asymmetric) scattering on defects of triangular shape. Introducing such defects to a regular honeycomb lattice violates the global spatial inversion symmetry (lowering the system symmetry to $C_{3v}$) and thus allows the appearance of ratchet flux. Within the framework of Schr\"odinger equation governing the spatiotemporal dynamics of exciton-polaritons in semiconductor microcavities, we provide detailed numerical simulations of wave packet scattering by a single defect of triangular shape. The obtained scattering cross sections and indicatrices highlight the microscopic details of skew scattering both for optical graphene and its classical counterpart.  Employing similar calculations we show the presence of macroscopic ratchet flux in the system of photonic graphene with multiple coherently oriented triangular defects. Finally, we support the obtained results with stochastic random walks approach drawing the correspondence between observed in simulations global ratchet effect and scattering cross-sections. As for the electronic ratchet effect in conventional graphene, optical ratchet effect is significantly more pronounced in classical regime of scattering. For triangle side $L$ greater than the wave length and wave packet size, classical wall reflection rules act and one can achieve maximal asymmetric scattering rate. Present study provides microscopic insight into skew scattering phenomenon in graphene and makes the concept of ratchet phenomena to be widen up on the field of photonics. The setup to experimentally measure the effect is proposed for the state-of-the-art etched microcavities with exciton-polaritons as well as other optical analogues of graphene.


%


%

\section*{acknowledgements}
We acknowledge the financial support from the Institute for Basic Science (IBS) in the Republic of Korea through the projects YSF project IBS-R024-Y3 and project IBS-R024-D1 (O.M.B.). We are thankful to V.V.~Koniakhin for proofreading and useful remarks.

\bibliography{references}






\setcounter{equation}{0}
\setcounter{figure}{0}
\setcounter{table}{0}
\setcounter{page}{1}
\setcounter{section}{0}

\makeatletter
\renewcommand{\theequation}{S\arabic{equation}}
\renewcommand{\thefigure}{S\arabic{figure}}
\renewcommand{\theHfigure}{S\arabic{figure}}
\renewcommand{\thetable}{S\arabic{table}}
\renewcommand{\thesection}{S\Roman{section}}
\renewcommand{\thepage}{S\arabic{page}}
\renewcommand{\bibnumfmt}[1]{[S#1]} 
\renewcommand{\citenumfont}[1]{S#1}

\onecolumngrid

\begin{center}
\textbf{\large Supplemental Material: Skew scattering and ratchet effect in photonic graphene}\\
\bigskip
O.M.~Bahrova$^{1,2,3}$, and S.V.~Koniakhin$^{1,4}$\\
\smallskip
\small{\textit{$^1$ Center for Theoretical Physics of Complex Systems,\\ Institute for Basic Science (IBS), Daejeon 34126, Republic of Korea \\
$^2$ B. Verkin Institute for Low Temperature Physics and
Engineering of the National\\ Academy of Sciences of Ukraine, 47
Nauky Ave., Kharkiv 61103, Ukraine\\
$^3$ Institut Pascal, PHOTON-N2, Universit\'e Clermont Auvergne,
CNRS, Clermont INP, F-63000 Clermont-Ferrand, France\\
$^4$ Basic Science Program, Korea University of Science and Technology (UST), Daejeon 34113, Republic of Korea}}

\end{center}

\onecolumngrid

\section*{List of supplemental figures}

\begin{figure}[!ht]
    \centering
    \includegraphics[width=0.99\linewidth]{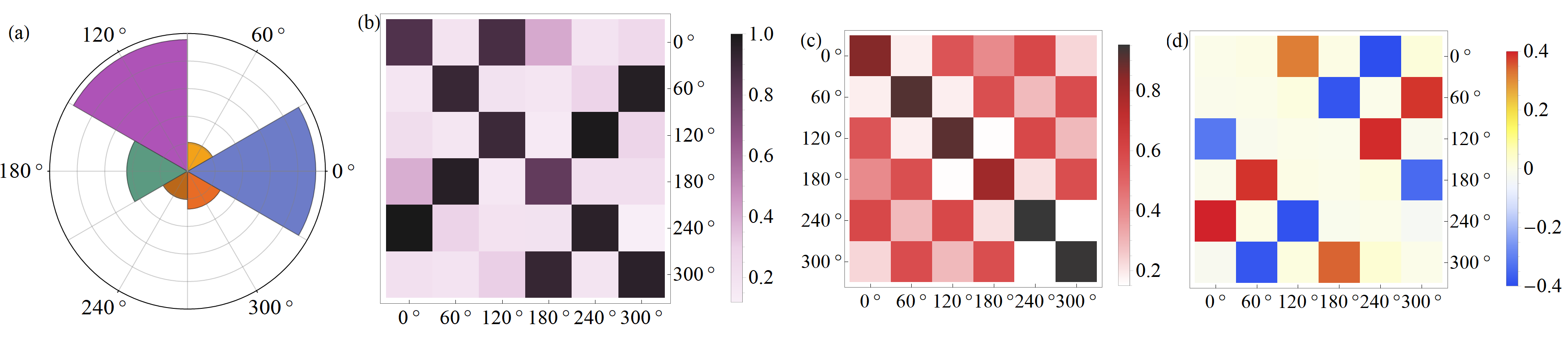}
    \caption{Real space scattering rate $W_{k,k'}$ in the "intermediate" regime normalized on the maximum (b). (a) scattering diagram $W_{0\degree,k'}$ representing in essence the first row of the matrix shown in (b). The symmetric (c) $W_{k,k'}^{(s)}$ and anti-symmetric (d) $W_{k,k'}^{(as)}$ parts of the scattering matrix $W_{k,k'}^{}$ (b).}
    \label{figS1}
\end{figure}

\begin{figure}[ht!]
    \centering
    \includegraphics[width=0.9\linewidth]{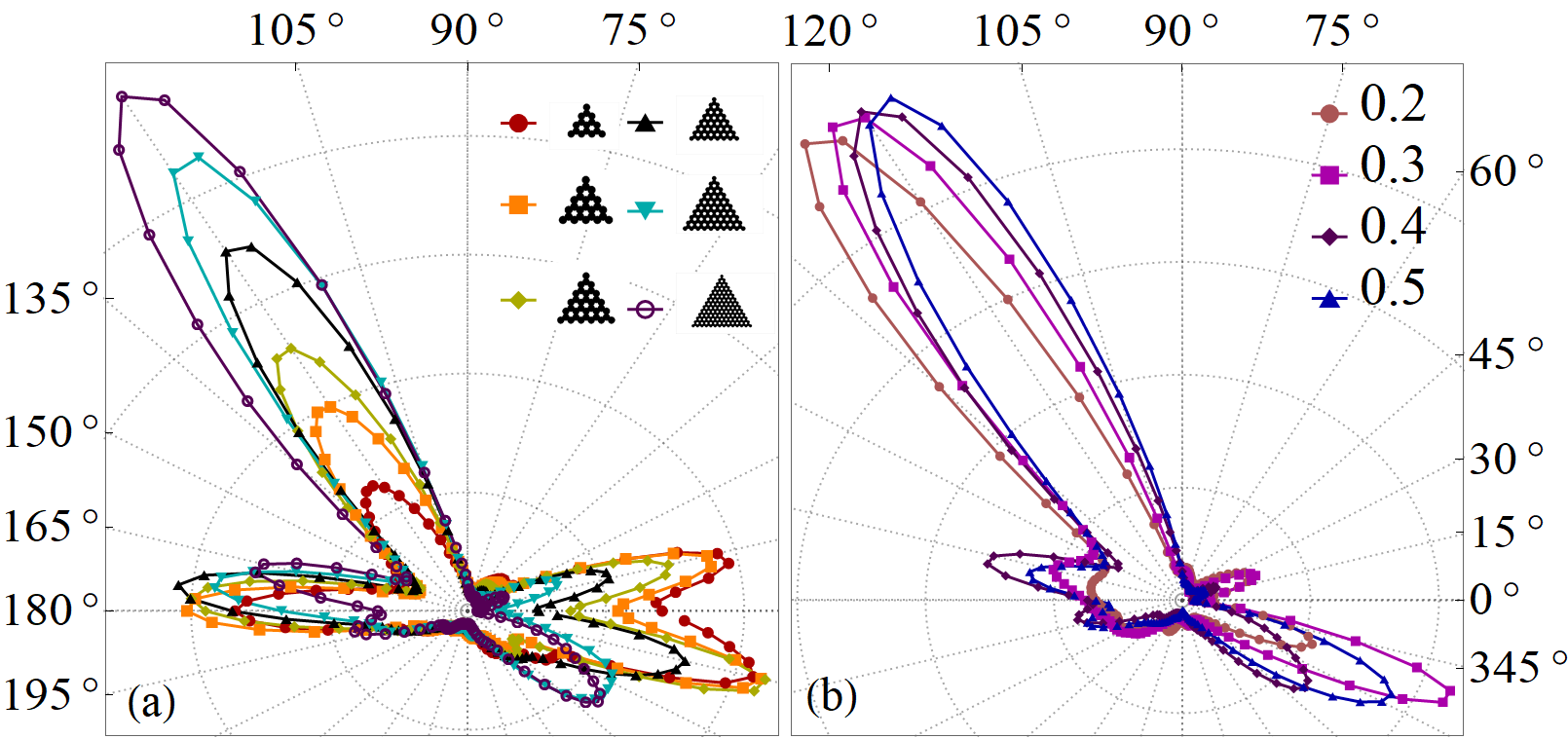}
    \caption{(a) Scattering diagrams for various sizes of triangle defect and $q=0.4$ (compare with the subfigure (a) in Fig.~2 of the main text). (b) Scattering diagrams for different values of initial wave vector for the regime of classical scattering. The purple curve is the same as one in (a).}
    \label{fig:figs2}
\end{figure}

\begin{figure}[ht!]
    \centering
    \includegraphics[width=0.5\linewidth]{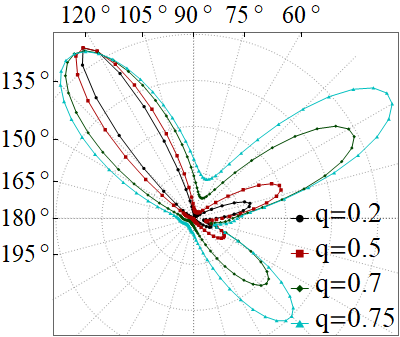}
    \caption{Scattering diagrams for different values of initial wave vector for the setup of scattering on a solid triangle. The green curve is the same as one in Fig.~2(b) of the main text.}
    \label{fig:figs3}
\end{figure}

\begin{figure}[ht!]
    \centering
    \includegraphics[width=0.99\linewidth]{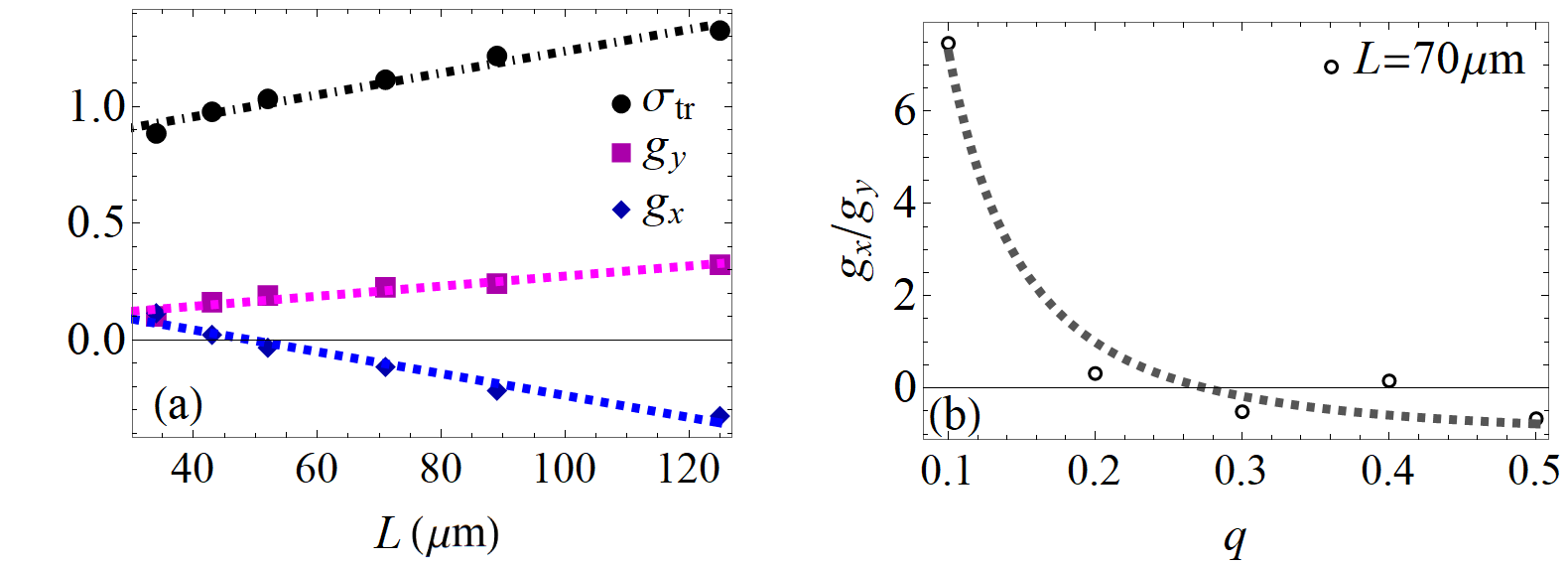}
    \caption{(a) Asymmetry parameter along the horizontal ($g_x$) and vertical ($g_y$) axes calculated for the scattering diagrams presented in Fig.~\ref{fig:figs2}(a). Additionally, black dots indicate the transport cross-section values $\sigma_{\text{tr}}$. See also plot (c) in Fig.~2 of the main text. (b) The ratio of $g_x/g_y$ versus initial wave vector values $q$ measured from the Dirac point one.}
    \label{fig:figs4}
\end{figure}

\begin{figure}[ht!]
    \centering
   \includegraphics[width=\linewidth]{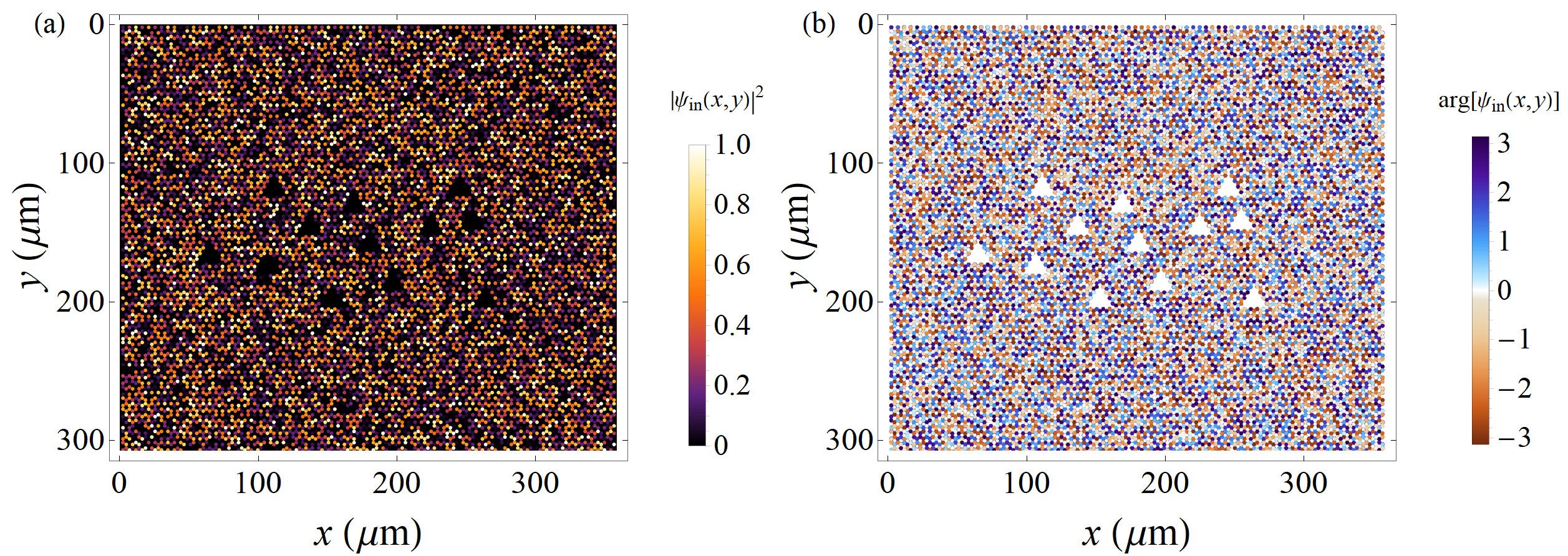}
   \caption{An example of random initial conditions used in numerical simulations of GPE in order to obtain the ratchet effect (see Fig.~5 of the main text). The initial wave function is chosen to be uniform random distribution for the amplitude (a) and phase (b) (a number is assigned for each lattice site (micropillar)). There is evidently no initially introduced preferential direction of motion. A sequence of prepared in a such way initial wave function with also randomly distributed triangle defects located in the central part of the sample, was used.  }
    \label{fig:figs5}
\end{figure}

\begin{figure}[ht!]
    \centering
    \includegraphics[width=0.99\linewidth]{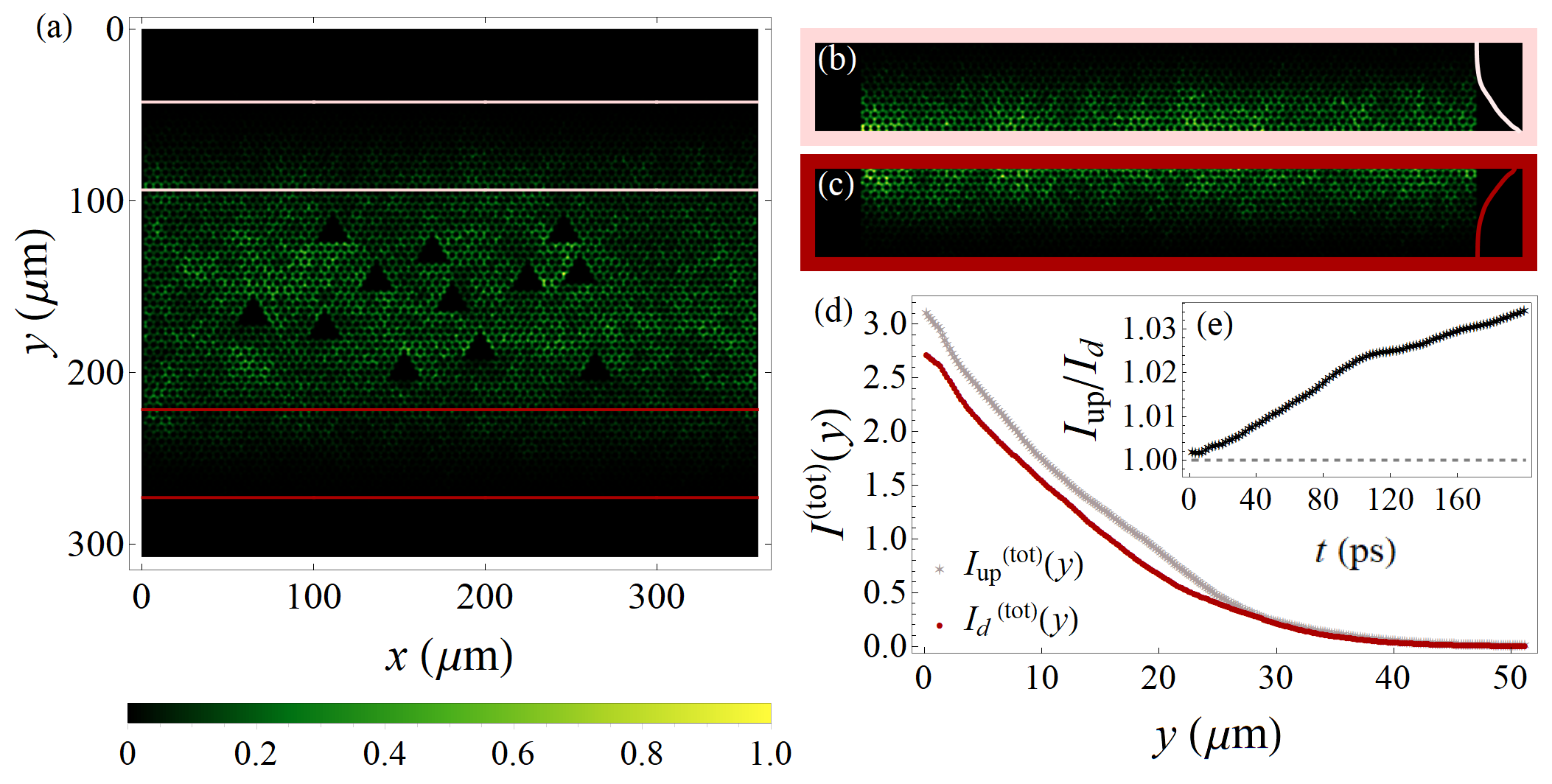}
    \caption{Time-averaged intensity (up to $t=190$ps). (e) The ratio of time-averaged wave function in upper region $I_{\mathrm{up}}$ to the one in lower counting region $I_\mathrm{d}$ versus time (averaged over $>$40 cases including shifts of the symmetry line of counting regions in respect to the defects center of mass and two different decay rates). In order to emphasize the difference in intensities, we plot $|\psi (x,y)|^6$ in (a). }
    \label{fig:figs6}
\end{figure}

\begin{itemize}
\addtolength{\itemindent}{0.33in}

    \item [{Figure~\ref{figS1}.}] Scattering diagram $W_{0\degree , k'}$ and scattering rate matrix for intermediate regime;
    \item [{Figure~\ref{fig:figs2}.}] Additional scattering diagrams (compare with Figs.~3-4 of the main text);
    \item [{Figure~\ref{fig:figs3}.}]  Scattering rate diagrams for the case of scattering on the solid triangle of the size as for Fig.~2(b);
    \item [{Figure~\ref{fig:figs4}.}] Dependencies of asymmetry vector components on the triangle size and the wave vector value for scattering diagrams Fig.~\ref{fig:figs2};
    \item [{Figure~\ref{fig:figs5}.}] An example of the wave function initial conditions used in 2D simulations of Schr\"odinger equation discussed in Sec.~III of the main text;
    \item [{Figure~\ref{fig:figs6}.}] Results of Schr\"odinger equation calculations analyzed in Sec.~III of the main text, demonstrating the emergence of ratchet effect.
\end{itemize}

%


%


\end{document}